\documentclass{jpp}

\usepackage{amsmath}
\usepackage{graphicx}
\usepackage{dcolumn}
\usepackage{wrapfig}
\usepackage{bm}
\usepackage{multirow}
\usepackage{xcolor} 
\usepackage{siunitx}
\usepackage{microtype}
\usepackage{hyperref}
\usepackage{multirow}
\hypersetup{
    linkcolor=blue,
    citecolor=blue,
    urlcolor=blue,
    colorlinks=true,
    pdfauthor={},
    pdftitle={}
}

\newcommand{\avg}[1]{\left\langle #1 \right\rangle}

\newcommand{\submax}{{\textup{max}}}
\newcommand{\alfven}{{\textup{A}}}
\newcommand{\plasm}{{\textup{p}}}
\newcommand{\collapse}{{\textup{col}}}
\newcommand{\bubble}{{\textup{bub}}}
\newcommand{\effective}{{\textup{eff}}}

\shorttitle{Beam current from downramp injection in electron-driven plasma wakefields}
\shortauthor{C. Hue, A. Golovanov, S. Tata, S. Corde, V. Malka}

\title{Beam current from downramp injection in electron-driven plasma wakefields}

\author{C\'eline Hue\aff{1}, Anton Golovanov\aff{1}\corresp{\email{anton.golovanov@weizmann.ac.il}}, Sheroy Tata\aff{1}, S\'ebastien Corde\aff{2} and Victor Malka\aff{1}}
\affiliation{\aff{1}Department of Physics of Complex Systems, Weizmann Institute of Science, 7610001 Rehovot, Israel
\aff{2}LOA, ENSTA Paris, CNRS, Ecole Polychnique, Institut Polytechnique de Paris, 91762 Palaiseau, France}
\date{\today}

\begin{document}

\maketitle

\begin{abstract}
We study the stability of plasma wake wave and the properties of density-downramp injection in an electron-driven plasma accelerator.
In this accelerator type, a short high-current electron bunch (generated by a conventional accelerator or a laser-wakefield acceleration stage) drives a strongly nonlinear plasma wake wave (blowout), and accelerated electrons are injected into it using a sharp density transition which leads to the elongation of the wake.
The accelerating structure remains highly stable until the moment some electrons of the driver reach almost zero energy, which corresponds to the best interaction length for optimal driver-to-plasma energy transfer efficiency.
For a particular driver, this efficiency can be optimized by choosing appropriate plasma density.
Studying the dependence of the current of the injected bunch on driver and plasma parameters, we show that it does not depend on the density downramp length as long as the condition for trapping is satisfied.
Most importantly, we find that the current of the injected bunch primarily depends on just one parameter which combines both the properties of the driver (its current and duration) and the plasma density.

\end{abstract}

\section{\label{sec:Intro}Introduction}

Plasma accelerators that rely on high-amplitude plasma wakefields are promising several orders of magnitude higher acceleration gradients than in conventional radio-frequency accelerators \citep{Malka_2008_NP_4_447, Esarey_2009_RMP_81_1229}.
The two main types of plasma accelerators are laser--wakefield accelerators (LWFAs) based on driving a plasma wake with a short intense laser pulse \citep{Tajima1979} and plasma--wakefield accelerators (PWFAs) based on using a short high-current particle bunch as a driver \citep{Chen_1985_PRL_54_693, Rosenzweig_1991_PRA_44_6189}.
LWFAs have demonstrated rapid growth both in terms of the energy of accelerated electrons (reaching the energies of 8\,GeV at the distance of 20\,cm \citep{Gonsalves_2019_PRL_122_84801}) and stability of beam properties \citep{Faure_2006_N_444_737}.
For PWFA, experimental studies have demonstrated high-gradient acceleration \citep{Blumenfeld_2007_Nature_445_741} and efficient energy transfer from the driver to the witness \citep{Litos_2014_Nat_515_92, Lindstroem_2021_PRL_126_14801}.

Despite some advantages such as the long dephasing length and a more stable wakefield structure, PWFAs saw comparatively less development than LWFAs because the sources of short (comparable to plasma wavelength) electron bunches with a high enough current to excite a nonlinear wake were not widely available.
Recently, the concept of hybrid LWFA--PWFA multi-staged plasma accelerators \citep{Hidding_2010_PRL_104_195002, Kurz_2021_NatComm_12_2895, LWFA_PWFA_Foerster} based on the idea of using electron bunches from the first LWFA stage to drive a second PWFA stage has started gaining popularity, broadening the possibilities for experimental PWFA studies.
LWFA-produced electron bunches are very short and high-current, so they naturally have excellent parameters to drive a highly-nonlinear (blowout) plasma wake.
The density-downramp injection technique also proved to be an effective way of injecting electrons in the PWFA stage \citep{XU2017,Beam_driven_SI_simu,MartinezdelaOssa_2017_PRAB_20_91301, Zhang_2019_PRAB_22_111301, Couperus2021Hybrid}. 
Even though the electron-driven second stage cannot significantly surpass the performance of the first LWFA stage in terms of the total energy of the accelerated electrons, due to its stable nature it can serve as a ``quality booster'' by generating bunches with improved properties \citep{MartinezdelaOssa_2017_PRAB_20_91301, LWFA_PWFA_Foerster}.
However, although relatively high energy transfer efficiency is achieved and high-quality beams with hundreds of MeVs are predicted, the energy stability for such accelerator in previous research is still comparable to the a single-stage LWFA accelerators \citep{LWFA_PWFA_Foerster}, and more studies are required to understand the parameter dependence and physics behind this stage.

For PWFA to be efficient, one of the important steps is to optimize the driver-to-witness energy transfer efficiency.
The beam currents of both the driver and the witness beams play decisive roles in optimizing the energy transfer efficiency.
Few studies so far have systematically focused on controlling the beam currents produced in plasma-based accelerators.
The production in LWFA of the current profile finely tuned for the use in the second stage was studied in \citet{MRE}, but studies for PWFAs are lacking.

In this paper, we focus on performance of PWFA based on density-downramp injection of electrons.
One important step in calculating and optimising the energy transfer efficiency is to understand the beam dynamics.
Previous research described such important phenomena as the hosing instability experienced at the tail of the bunch \citep{hosing_huang_2007}, beam head erosion \citep{Zhou_2007_HeadErosion,Li_2012_HeadErosion}, energy depletion \citep{Muggli_2010_EnergyDepletion}, and the transformer ratio \citep{Blumenfeld_TransferRatio_2010}.
Yet, driver parameters used in these studies corresponded mostly to electron beams produced by conventional linear accelerators, which are very different from beams produced in LWFA stage for hybrid accelerators.
In this article, the dynamics of flattop--current electron drivers with a total beam charge of 100--500 pC, 100--300 MeV energy typical for the LWFA-produced bunches \citep{MRE} is studied.
In Section \ref{sec: Energy depletion}, the stability of the wakefield and driver-to-plasma energy transfer efficiency are discussed.
In Section \ref{sec:PWFA SI}, we numerically study density-downramp injection in the PWFA stage and the dependence of the current of the injected electron bunch on the parameters of the driver, the plasma, and the downramp.
Despite having such a multidimensional parameter space, we demonstrate that the injected current is determined mostly by one parameter, the effective current $J_\effective = J_b (k_\plasm \xi_b)^{2/3}$, which combines both the properties of the driver (its current $J_b$ and length $\xi_b$) and the plasma (wavenumber $k_\plasm$).
The dependence of the witness current on this parameter is linear.
We also show the limitation of this scaling for longer driver bunches which cannot efficiently excite a nonlinear wake.

\section{\label{sec: Energy depletion}Driver stability and driver-to-plasma efficiency}

In this section, we study the evolution and propagation of an electron beam in PWFA assuming fully pre-ionized plasma.
For sufficiently short, tightly-focussed and high-current beams, the wakefield is excited in the strongly nonlinear (``bubble'' or ``blowout'') regime \citep{Rosenzweig_1991_PRA_44_6189, Pukhov_2002_APB_74_355, Lotov_2004_PRE_69_46405}, when a cavity (a bubble) devoid of plasma electrons is formed behind the driver.
The excitation of the bubble by an electron beam can be self-consistently described by a model by \citet{Golovanov_2021_PPCF_63_85004} which can be used to calculate the shape of the bubble as well as the distribution of all the fields in it based solely on the charge density distribution of the driver.
As this model is based on the relativistic limit of the theory by \citet{Lu2006}, it is strictly valid only in the case when the transverse size of the bubble $R_\bubble$ is large in terms of plasma units, $k_\plasm R_\bubble \gg 1$, which is not necessarily the case for the parameters considered in the paper.
A more accurate model for comparatively small bubbles ($k_\plasm R_\bubble \sim 1$) was recently proposed in \citet{Golovanov_2023_PRL}, but it lacks simple analytical scalings.
Theoretical models also cannot fully describe the driver evolution and self-injection, and thus cannot completely replace numerical simulations.
Still, we can use the predictions of the models to compare to the simulation results.

\begin{figure}
    \centering
    \includegraphics[width=\linewidth]{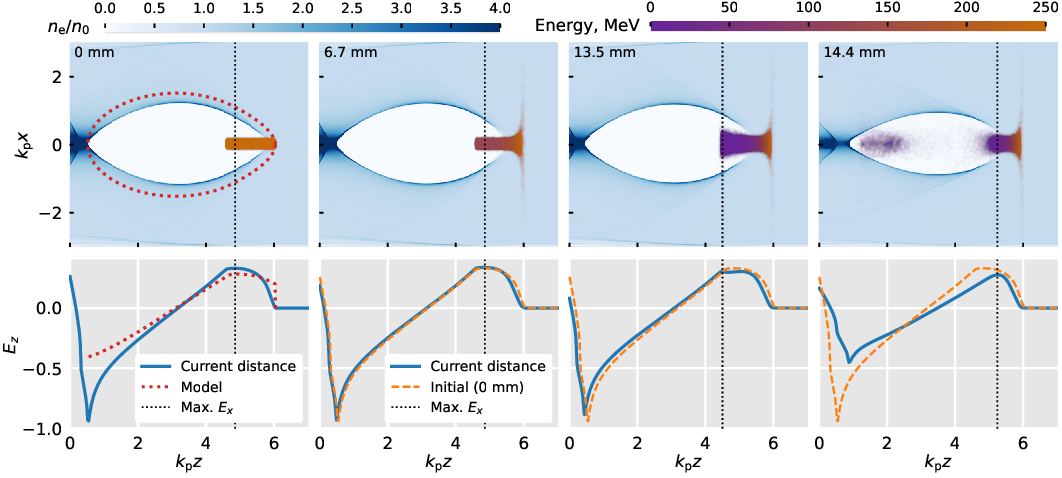}
    \caption{Electron density $n_\mathrm{e}$ distributions and the longitudinal electric fields $E_z$ on the axis in a wake driven by an electron driver at different propagation distances.
    The dashed lines show the initial distribution of the accelerating field at 0\,mm.
    For the driver, the color shows the energy of particles.
    The dotted lines at 0\,mm show the analytical solution according to the model in \citet{Golovanov_2023_PRL}.
    Vertical dotted lines show the position of the maximum decelerating field $E_z$.
    Plasma density $n_0 = \SI{3.125e17}{cm^{-3}}$.}
    \label{fig:electron_driver_collapse}
\end{figure}

To study the evolution of the driver in PWFA, we perform numerical simulations of the beam--plasma interaction using the 3D quasistatic particle-in-cell (PIC) code QuickPIC \citep{Huang_2006_JCP_217_658}.
The beam has the mean energy of \SI{250}{\MeV} and the charge of \SI{137}{pC} (corresponding to the total energy of \SI{17}{mJ}) with a flat-top longitudinal current profile with the length $\xi_b = \SI{13.4}{\um}$ (or the duration \SI{45}{fs} corresponding to the current of \SI{3.1}{kA}) and a Gaussian transverse profile $\exp(-r^2/2\sigma_r^2)$ with the radius $\sigma_r = \SI{0.52}{\um}$.
The peak number density of the beam $n_b = \SI{3.75e19}{\cm^{-3}}$.
The chosen parameters correspond to typical electron beams generated by the density-downramp injection in the first LWFA stage and are taken from simulations in \citet{MRE}. 
In fact, as will be briefly explained in the next section, the tunability of the LWFA-produced driver parameters is fairly limited.

The plasma density $n_0$ is chosen to be much lower than the density of the beam $n_b$, so that the driver excites a highly non-linear plasma wakefield, or a bubble (see Fig.~\ref{fig:electron_driver_collapse}).
The transverse emittance of the beam is chosen to be close to matched to the plasma density to prevent significant changes of the transverse size.
In this wakefield, the driver generally experiences two forces: the focussing force from the ion column (linear in the distance $r$ from the axis inside the bubble and leading to betatron oscillations of the electrons of the driver) and the decelerating longitudinal force.
We observe that the accelerating structure remains highly stable before rapidly collapsing when electrons of the driver start dephasing due to deceleration to very low energies.
This inherent stability owes to the fact that the transverse distribution of the matched driver does not influence the structure of the bubble, while the longitudinal velocity of the ultrarelativistic particles of the bunch stays almost equal to the speed of light until some of the electrons are decelerated to sub-relativistic energies, corresponding to the moment of collapse (compare the first three distances in Fig.~\ref{fig:electron_driver_collapse} before the collapse to the fourth one after the collapse began).

The propagation length corresponding to this collapse is much shorter than the distance of particle loss due to head erosion or beam hosing instability, so it effectively determines the PWFA stage length.
Head erosion happens due to the initial emittance of the driver in the region at the head of the bunch where the wakefield providing the focusing force is not high enough amplitude yet \citep{Zhou_2007_HeadErosion,Li_2012_HeadErosion}.
The Coulomb self-force is proportional to $\gamma^{-2}$ and can be usually ignored for ultrarelativistic bunches.
In the blowout regime of PWFA, head erosion affects only a small portion of the bunch at the head, as the wakefield amplitude quickly becomes sufficient to maintain focusing of almost the entire bunch.
As the head erodes, it still continues to drive a lower-amplitude wake behind it, which makes the spread of this erosion to other parts of the bunch very slow.
As can be seen from Fig.~\ref{fig:electron_driver_collapse}, it leads only to a small shift in the phase of the nonlinear wake over the propagation distance and cannot significantly affect the acceleration process.

Another effect which is believed to limit the length of PWFAs is the hosing instability which should lead to the exponential growth in oscillations of the beam centroid and the corresponding growth of oscillations of the bubble \citep{hosing_huang_2007}.
However, recent research suggests that energy depletion of the driver as well as its energy spread or energy chirp can efficiently suppress and saturate this instability due to the dephasing of betatron oscillations \citep{Mehrling_2017_PRL_118_174801, Mehrling_2019_PRAB_22_31302}.
In our simulations, hosing is not taken into account as the initial charge density distribution is ideally symmetric.

So, energy depletion of the driver is the main mechanism which determines the acceleration length.
It leads to electrons reaching sub-relativistic energies, after which significant portion of the electrons of the driver are quickly lost and the accelerating structure collapses (see Fig.~\ref{fig:electron_driver_collapse} at 14.4\,mm).
For a bunch with no energy chirp, this process happens at the point of the peak decelerating electric field the value of which can be estimated from the solution based on the model in \citet{Golovanov_2023_PRL} (shown with dotted lines in Fig.~\ref{fig:electron_driver_collapse} at 0\,mm).
The length at which the accelerating structure collapse happens can be thus estimated as the length at which an electron with the kinetic energy $K \approx \gamma m c^2$ is decelerated to zero energy in the field $E_\submax$,
\begin{equation}
    L_\collapse \approx \frac{\gamma m c^2}{e E_\submax},
    \label{eq:L_collapse}
\end{equation}
where $\gamma$ is the Lorentz factor, $m$ is the electron mass, $e > 0$ is the elementary charge, and $c$ is the speed of light.
The comparison to values observed in the simulations (see Table~\ref{tab:driver_collapse}) provides a fairly good estimate for the collapse length.

\begin{table}
    \centering
    \begin{tabular}{r  r r | r r r r}
          & & & \multicolumn{2}{c}{$L_\collapse[\si{mm}]$} & \multicolumn{2}{c}{$\eta$, \% }  \\
         $n_0[\si{cm^{-3}}]$ & $n_b/n_0$ & $k_\plasm \xi_b$ & sim. & model & sim. & model \\ \hline
         \num{2.5e18} & 15 & 3.99 & 4.8 & 6.0 & 65 & 73 \\
         \num{1.875e18} & 20 & 3.45 & 5.5 & 7.1 & 71 & 81 \\
         \num{1.25e18} & 30 & 2.82 & 6.7 & 8.1 & 76 & 87 \\
         \num{6.25e17} & 60 & 1.99 & 9.4 & 11.6 & 77 & 93 \\
         \num{4.17e17} & 90 & 1.63 & 11.6 & 14.1 & 75 & 93 \\
         \num{3.125e17} & 120 & 1.41 & 13.5 & 16.4 & 73 & 92 \\
    \end{tabular}
    \caption{The collapse length of the electron driver $L_\collapse$ observed in quasistatic PIC simulations and estimated from Eq.~\eqref{eq:L_collapse} as well as the efficiency $\eta$ calculated according to Eq.~\eqref{eq:driver_efficiency} and estimated based on the model from \citet{Golovanov_2023_PRL} for different plasma densities $n_0$.}
    \label{tab:driver_collapse}
\end{table}

When the collapse of the accelerating structure begins, the driver still has part of its energy left (as decelerating field is non-uniform, and different parts of the driver experience different field values), which limits driver-to-plasma energy transfer efficiency defined as the percentage of the bunch initial energy spent on generating the wakefield by the moment the collapse begins,
\begin{equation}
    \eta = \frac{\avg{\gamma}_0 - \avg{\gamma}_\collapse}{\avg{\gamma}_0},
    \label{eq:driver_efficiency}
\end{equation}
where averaging is performed over the bunch particles.
Assuming that the bunch is monoenergetic, it can also be estimated as $\eta \approx \avg{E_z}/E_{\submax}$, so the efficiency mostly reflects how uniform the distribution of the decelerating electric field inside the driver is.
The comparison between the actual efficiency observed in simulations to the estimated efficiency based on the field distribution $E_z$ according to the model by \citet{Golovanov_2023_PRL} is shown in Table~\ref{tab:driver_collapse}.
The efficiency changes depending on the plasma density, so it is required to carefully choose the plasma density for the second PWFA stage in order to increase the energy transfer efficiency.
For a very high plasma density, the driving bunch can become too long compared to the plasma wavelength (determined by the dimensionless value $k_\plasm \xi_b$ in Table~\ref{tab:driver_collapse}), leading to its tail being accelerated in the created bubble and a significant drop in the driver-to-plasma energy transfer efficiency due to non-uniformity of the field distribution.
The efficiency can thus be increased by lowering the plasma density.
In very low-density plasmas, the efficiency also starts to slightly drop due to the non-uniformity of the field at the front of a short driver.
In addition to that, lower densities can be less desirable due to lower acceleration gradients, which increase the size of the accelerator.
Therefore, there is an optimal range of plasma densities at which the beam-to-plasma efficiency is high.
As results of numerical simulations show (see Table~\ref{tab:driver_collapse}), for the considered electron driver, the optimal plasma density of the second PWFA stage is around \SI{e18}{\cm^{-3}}, corresponding to the length of the driver in plasma units $k_\plasm \xi_b \sim 3$.

\section{\label{sec:PWFA SI} Downramp injection}

\begin{figure}
\includegraphics[]{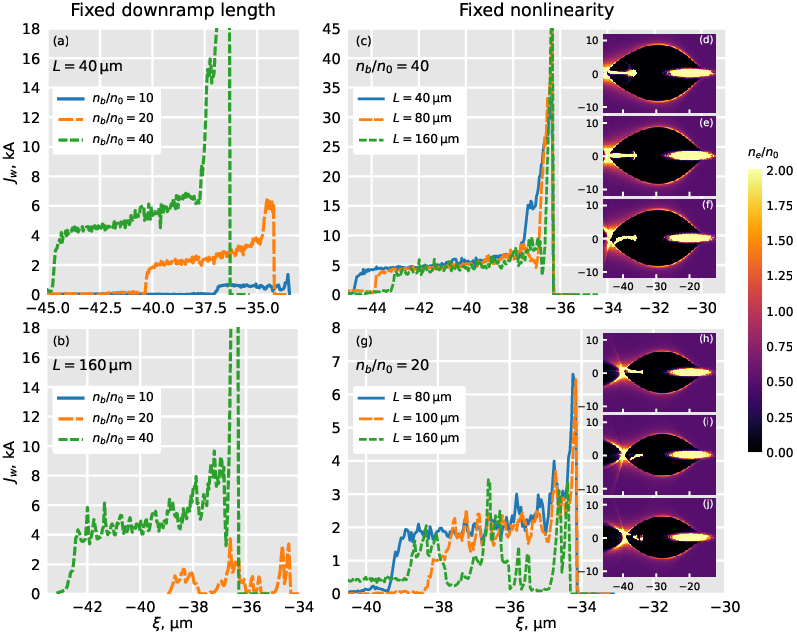}
\caption{\label{fig:scan_downramp_L} Current profiles of injected beam $J_w$ corresponding to different lengths of plasma density downramp $L$ and different wakefield non-linearity $n_b/n_0$. The inset sub-figures are wakefield structures for the plotted simulations. Plasma density before the downramp $n_0 = \SI{5e18}{\cm^{-3}}$; $n_0/2$ after the downramp. The driver has the Gaussian shape with $\sigma_z = \SI{2.5}{\um}$, $\sigma_r = \SI{0.5}{\um}$.} 
\end{figure}

In this section, we study the dependence of the current profile of the witness electron beam produced by density-downramp injection on the driver and plasma parameters.
A recent study on downramp injection for LWFA \cite{MRE} reports that the injected beam current is strongly influenced by the laser intensity, and, consecutively, the wakefield strength.
Unlike for the laser driven counterpart, the electron driver can be considered non-evolving during the beam injection process in the PWFA stage since the beam evolution scale is the period of betatron oscillations $\sqrt{2 \gamma} \lambda_\plasm$, usually of millimeter to centimeter scale, much longer than the downramp length, usually in the range of hundreds of micrometers or less. As a result, the location of the downramp does not play such a crucial role as in the LWFA case.

\subsection{\label{sec:Downramp_steepness}The influence of downramp steepness}

Now we consider the influence of downramp steepness on the parameters of the injected beam.
As quasistatic codes like QuickPIC used in Sec.~\ref{sec: Energy depletion} cannot self-consistently describe the injection of particles into the wakefield, we perform PIC simulations with FBPIC \citep{Lehe_2016_CPC_203_66}.
The downramp is modelled as a linear change of plasma density from $n_0$ to $n_0/2$ of length $L$.
For a given downramp steepness, particles are injected only when their energy is high enough, the required energy grows with the decrease of the downramp steepness \citep{XU2017}.
The energy of the particles depends only on the nonlinearity of the created bubble (which will be quantified later) irrespective of other driver properties.

Four group of simulations are presented in Fig.~\ref{fig:scan_downramp_L} that explain the role that the non-linearity of the bubble (which depends on the density ratio $n_b/n_0$ between the driver and the plasma for a fixed driver shape and size) and $L$ play on the witness injection.
The injected beam currents are plotted and the corresponding wakefield structures are illustrated in the inset of subfigures (c) and (g).
As subfigure (b) demonstrates, for a fixed downramp length $L$, the injection occurs only when the bubble non-linearity is strong enough.
Also, once the condition for injection is reached, the witness beam current is barely influenced by the downramp steepness for the same bubble strength (the same driver), as shown in subfigure (c).
Similar to the results in \citet{MRE}, a current peak is observed at the head of the witness bunch.
This happens due to the non-linear phase mixing caused by a sharp plasma density transition when the downramp begins, and the peak can be mitigated with a smooth density transition at the beginning of the downramp, usually found in experimental cases.
Behind the head of the bunch with respect to the co-moving coordinate $\xi = z - ct$, the injected current stabilises at a constant value.
In the following considerations, we will determine the value of the injected witness current $J_w$ as the current of the constant part of the profile.
As subfigures (c, g) and corresponding inset figures show, the length and thus the charge of the injected beam also does not depend much on the transition length $L$.
It is mostly determined by the ratio between the initial and the final plasma density (fixed at 2 in our case), as this ratio determines the change of the length of the bubble during the transition.

As shown by Fig.~\ref{fig:scan_downramp_L}(a) and (b), the higher value of ratio $n_b/n_0$ (and thus the wakefield non-linearity) leads to the higher current $J_w$.
From the comparison of subfigures (a) and (b), one can see that shorter and steeper downramp enables injection at lower driver charge and lower non-linearity.
No injection at all is observed for the case where $n_b/n_0=10$ shown in sub-figure (b), and the higher the wakefield non-linearity $n_b/n_0$ is, the stronger the injected witness current becomes.

\subsection{Injected beam current scaling}
 
As shown in the previous section, as long as the density downramp provides stable injection, its properties do not significantly affect the injected current $J_w$, so it should mostly be determined by the parameters of the driver and their relation to the plasma density.
It is suggested by Fig.~\ref{fig:scan_downramp_L} that the higher values of $n_b/n_0$ which should correspond to stronger nonlinearity lead to a stronger injected beam current.
As we want to investigate the dependence of $J_w$ on the nonlinearity of the bubble, we need to introduce a measure of it first.
In the previous section, we used the ratio $n_b/n_0$ as this measure, which is only suitable for a fixed shape and size of the driver.
A more general measure requires introducing the energy properties of the bubble.

Following \citet{Lotov_2004_PRE_69_46405, Golovanov_2021_PPCF_63_85004, Golovanov_2023_PRL}, we introduce the quantity $\Psi(\xi) = \int (c W - S_z) d^2r_\perp$ which depends on the comoving coordinate $\xi = ct - z$ and contains the integral over the transverse plane of the energy density $W$ and the longitudinal energy flux $S_z$ of both the EM field and the plasma particles.
The value of $\Psi$ is also equal to the total energy flux in the comoving window \citep{Lotov_2004_PRE_69_46405}.
As it has the dimension of power, we will refer to $\Psi$ as ``the power of the bubble''.
In the absence of energy exchange with bunches, $\Psi$ is a conserved property in any wake; drivers lead to the increase of $\Psi$, while accelerated witnesses decrease it.
As a property describing the energetic properties of the bubble, it is equal to the total power of deceleration felt by the driving bunch and is also equal to the maximum achievable power of acceleration for the witness bunch.
For the blowout regime of plasma wakefield, $\Psi$ is fully determined by the size of the bubble $R_\bubble$ and grows with it; in the limit of a large bubble size ($k_\plasm R_\bubble \gg 1$), $\Psi \propto (k_\plasm R_\bubble)^4$ (see \citet{Tzoufras_2008_PRL_101_145002, Golovanov_2021_PPCF_63_85004}).
The power of the bubble $\Psi$ serves as quantification of nonlinearity of the wake and it is the most important property of the bubble.
Regardless of the shape the driver which creates a bubble with a certain value of $\Psi$, the effect of the bubbles with the same power on acceleration of particles and on downramp injection will be mostly the same.
Therefore, we can expect that injection only depends on the value of $\Psi$.

According to \citet{Golovanov_2021_PPCF_63_85004}, for a blow-out wakefield excited by a sufficiently tighly focussed ($k_\plasm r_b \ll 1$) driver with the charge $Q$ and a flattop current profile of length $\xi_b$, the power of the bubble in the large-bubble limit ($k_\plasm R_\bubble \gg 1$) is calculated using the following formula:
\begin{equation}
    \Psi \approx \frac{4\pi m^2 c^6 \varepsilon_0}{e^2 J_\alfven} k_\plasm Q \left[\sqrt{\frac{2 c Q}{J_\alfven \xi_b}} - \frac{k_\plasm \xi_b}{8} \right],
\end{equation}
where $J_\alfven = 4 \pi \varepsilon_0 m c^3 / e \approx \SI{17}{kA}$ is the Alfv\'{e}n current, $\varepsilon_0$ is the vacuum permittivity.
It can be rewritten as
\begin{equation}
    \Psi \approx \frac{\sqrt{2} m c^2 J_\alfven}{e} \left(\frac{J_\effective}{J_\alfven}\right)^{3/2} \left[1 - \frac{(k_\plasm \xi_b)^{4/3}}{\sqrt{128 J_\effective / J_\alfven}} \right],
    \label{eq:Psi_formula}
\end{equation}
where we introduce the quantity $J_\effective$ which we call ``the effective current'' of the driver:
\begin{equation}
    J_\effective = \frac{c Q}{\xi_b} (k_\plasm \xi_b)^{2/3} = J_b (k_\plasm \xi_b)^{2/3}.
    \label{eq:Jeff}
\end{equation}
For high-current ($J_\effective \sim J_\alfven$) sufficiently short electron bunches, the last factor in Eq.~\eqref{eq:Psi_formula} is close to $1$, and $\Psi \propto J_\effective^{3/2}$.
In general, this factor describes the weakening of the bubble when the bunch becomes long enough compared to the plasma wavelength.
In the limiting case when the back of the bunch is already located in the accelerating phase, the bubble effectively transfers the energy from one part of the driver to another, limiting the possible efficiency of accelerating the witness, which corresponds to this factor tending to 0.
When it becomes negative, this solution for $\Psi$ is formally incorrect, which corresponds to the situation when the driver cannot fit inside the bubble.

Since $J_\effective$ fully determines $\Psi$ and thus the properties of the bubble for sufficiently short bunches, it should be the only combination of the driver and plasma parameters affecting the injection process.
Although the model in \citet{Golovanov_2021_PPCF_63_85004} is not necessarily strictly valid in the case of smaller-size bubbles, because this dependence holds in the important limit of large bubbles, we make a conjecture that $J_\effective$ should be the most important combination of the driver's parameters.

\begin{figure}
\includegraphics[width=\textwidth]{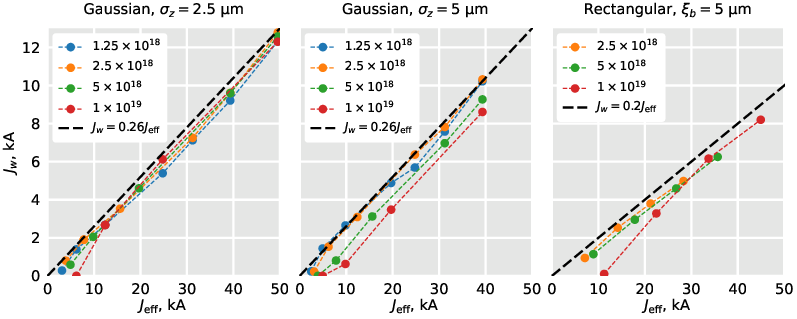}
\caption{\label{fig:Jw_const_length} The relationship between $J_w$ and $J_\effective$ for different types of beams. Each line corresponds to the same value of plasma density (in \si{cm^{-3}} in the legend), and the driver beam current $J_b$ is varied while the beam longitudinal size ($\sigma_z$ or $\xi_b$) remains the same.
The dashed black line corresponds to the linear dependence on $J_w$.}
\end{figure}

\begin{figure}
\includegraphics[width=\textwidth]{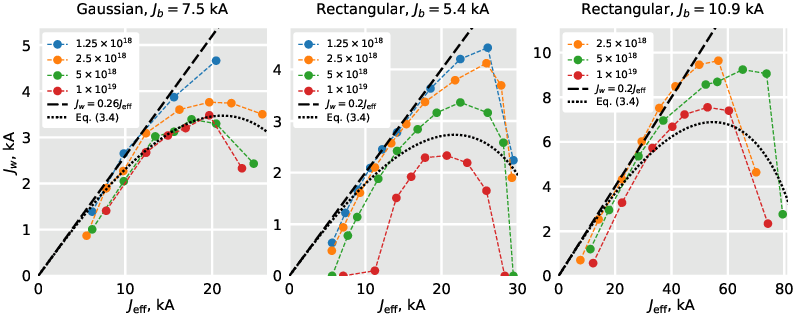}
\caption{\label{fig:Jw_const_current} The relationship between $J_w$ and $J_\effective$ for different types of beams and different beam currents. Each dashed line corresponds to the same value of plasma density (in \si{cm^{-3}} in the legend), and the driver beam longitudinal size $\xi_b$ is varied while the driver current $J_b$ remains the same.
The dashed black line corresponds to the linear dependence on $J_w$.
The dotted line shows the prediction according to Eq.~\eqref{eq:Jeff_corrected} to with a correction $c_2 = 0.8$ for the Gaussian shape and $0.4$ for the rectangular shape.}
\end{figure}

To explore the influence of $J_\effective$ on the injected witness current $J_w$, we perform numerical PIC simulations with FBPIC \citep{Lehe_2016_CPC_203_66} for two types of drivers: 3D Gaussian beams which are usually used to model the driver beam produced by a conventional accelerator \citep{FACET_II_Joshi_2018,FLASH_Forward_2019}, and flattop longitudinal bunch which models the LWFA produced driver beam \citep{Couperus2021Hybrid, LWFA_PWFA_Foerster}. To prevent any effects due to the evolution of the driver beam, we artificially freeze it by setting its initial particle energy to \SI{50}{GeV}. The ratio between the plasma density before and after the downramp is fixed to 2 (going from $n_0$ to $n_0/2$) and the downramp length is between \SI{20}{\um} and \SI{160}{\um}. For Gaussian drivers with the density profile defined as $n_b \exp[-r^2 / 2 \sigma_r^2 - (\xi-\xi_0)^2 / 2 \sigma_z^2]$, the beam transverse size $\sigma_r = \SI{0.5}{\um}$. Flattop drivers correspond to the cylinder of constant density $n_b$, length $\xi_b$, and fixed radius $r_b = \SI{0.6}{\um}$.

As the effective current $J_\effective$ defined by Eq.~\eqref{eq:Jeff} depends on the current of the driver $J_b$ and its length $\xi_b$, we study the relationship $J_w$ and $J_\effective$ by varying these two parameters.
Of course, the value of $J_\effective$ also depends on the plasma density, so we use the plasma density $n_0$ before the downramp to calculate the plasma wavenumber $k_\plasm$ used in Eq.~\eqref{eq:Jeff}.
Because definition Eq.~\eqref{eq:Jeff} is written for flattop profiles, for Gaussian beams we use $\xi_b = \sqrt{2} \sigma_z$ and the peak value of $J_b$.
This combination was shown to provide a good approximation for $\Psi$ in simulations (not presented in the paper).
The results for varying the driver beam current $J_b$ at fixed length are shown in Fig.~\ref{fig:Jw_const_length}, and the results for varying the driver bunch length while the current is fixed are presented in Fig.~\ref{fig:Jw_const_current} for different plasma densities.

For the fixed beam length (Fig.~\ref{fig:Jw_const_length}), we observe a linear dependence of the injected current on the effective current, which seems to indicate that $J_w \propto \Psi^{2/3} \propto J_\effective$.
The slope of the linear dependence depends only on the shape of the driver and remains the same for different plasma densities and lengths of the driver. 
We observe that the witness current becomes slightly lower for denser plasmas (thus larger values of $k_\plasm$), which corresponds to the lowering efficiency due to the driver length in plasma units, as predicted by the length factor in Eq.~\eqref{eq:Psi_formula}.
At very low $J_\effective$, the injected current goes to 0, which corresponds to the injection threshold discussed in Sec.~\ref{sec:Downramp_steepness}.

\begin{figure}
\centering
\includegraphics[]{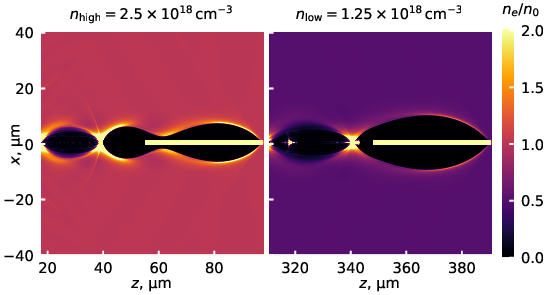}
\caption{\label{fig:Wakefield long bunch} Electron density distribution in a wake excited by a long rectangular-shape driver before and after the downramp.
The downramp provides transition from $n_0$ to $n_0/2$ where $n_0 = \SI{2.5e18}{\cm^{-3}}$.
The parameters of the driver: $n_b = 40 n_0$, $\xi_b = \SI{42.5}{\um}$, $r_b = \SI{0.6}{\um}$.}
\end{figure}

To study the dependence of the witness current $J_w$ on the driver length $\xi_b$ in more details, we also plot the dependence of $J_w$ on the effective current $J_\effective$ for a fixed current of the driver, which means that the increase in $J_\effective$ corresponds to the increase in its length (Fig.~\ref{fig:Jw_const_current}).
For smaller $J_\effective$ and thus shorter driver lengths, the already found linear trend with the same slope is again observed for both types of drivers.
But for higher $J_\effective$ and longer driver length, we see the deviation of $J_w$ from the linear trend and the decline in it.
This behavior is qualitatively consistent with Eq.~\eqref{eq:Psi_formula} which predicts that the elongation of the driver lowers the efficiency of exciting a bubble, so we can expect that the witness current $J_w \propto \Psi^{2/3}$ behaves like
\begin{equation}
    J_w = c_1 J_\effective \left[1 - c_2 \frac{(k_\plasm \xi_b)^{4/3}}{\sqrt{128 J_\effective/J_\alfven}} \right]^{2/3}.
    \label{eq:Jeff_corrected}
\end{equation}
By choosing the value of $c_2$, the qualitative behavior of $J_\effective$ can be recreated using this formula (see dotted lines in Fig.~\ref{fig:Jw_const_current}).

However, this formula cannot show the full picture, because downramp injection for longer drivers is more complex than for short ones.
According to Eq.~\eqref{eq:Psi_formula}, the power of the bubble changes as the driver propagates through the density downramp which changes the plasma wavenumber $k_\plasm$.
For short drivers, $\Psi \propto J_\effective^{3/2} \propto n_\plasm^{1/2}$ scales exactly the same for all drivers with the same $J_\effective$, and thus the injected current is still fully determined by $J_\effective$.
However, longer drivers which have a lowered efficiency in dense plasma begin exciting the bubble more effectively when transitioning into lower density plasma at the downramp, as their length in plasma units becomes small.
An example of such a driver is shown in Fig.~\ref{fig:Wakefield long bunch}: in the initial larger density it is too long to fit inside the bubble, so during the density downramp it passes through the point of having almost zero efficiency of exciting a bubble ($\Psi \approx 0$), and then $\Psi$ starts growing again as the driver becomes shorter than the excited bubble.
In addition to the complex behaviour of the nonlinearity of the bubble, the corresponding velocity of the back of the bubble which determines the injection threshold will also significantly depend on the length of the driver.
The dynamics of downramp injection for such cases cannot be reduced to a simple formula such as linear dependence on $J_\effective$ or even corrected Eq.~\eqref{eq:Jeff_corrected} and requires the full description of the evolution of the bubble in the downramp.
However, as explained in Sec.~\ref{sec: Energy depletion}, these cases are suboptimal for efficient utilisation of the driver's energy and should be avoided.
In more optimal cases, when the driver is comparatively short, the linear dependence of $J_w$ and $J_\effective$ holds.


\section{Conclusion}

We studied a plasma-wakefield accelerator driven by a short high-current electron bunch.
The accelerating structure in this case remains stable until the moment when some of the electrons of the driver lose all their energy, which is the condition which determines the acceleration length optimizing the driver-to-plasma energy transfer efficiency.
The dependence of the witness bunch generated by density-downramp injection on the parameters of the driver was studied.
We showed that a steeper downramp enables injection for a weaker driver, but the current of the injected bunch does not significantly depend on the downramp length as long as the criterion for the injection is met.
The witness current of the injected bunch mostly depends on the effective current of the driver $J_\effective$ which combines both the driver's current and its length in plasma units.
The dependence of the injected witness current on the effective current is linear for reasonably short drivers. 

This work was supported by the Fondation Jacques Toledano, the Schwartz Reisman Center for Intense Laser Physics, and by ERC PoC Vherapy grant.

\bibliographystyle{jpp}
\bibliography{apssamp2}

\end{document}